\documentclass[conference,10pt]{IEEEtran}
\IEEEoverridecommandlockouts
\usepackage{amsmath,amssymb,amsfonts}
\usepackage{algorithmic}
\usepackage{graphicx}
\usepackage{textcomp}
\usepackage[table,xcdraw]{xcolor}
\usepackage{xcolor}
\usepackage{todonotes}
\usepackage{caption}
\usepackage{subcaption}
\usepackage{hyperref}
\usepackage{float}
\usepackage[subtle,tracking=normal,bibnotes,charwidths]{savetrees}

\begin{document}

\title{The Efficacy of Transformer-based Adversarial Attacks in Security Domains\\
}

\author{}
\author{
    \IEEEauthorblockN{Kunyang Li, Kyle Domico, Jean-Charles Noirot Ferrand, Patrick McDaniel}
    \IEEEauthorblockA{University of Wisconsin-Madison
    \\\ \{kli253, domico, jcnf, mcdaniel\}@cs.wisc.edu}
}

\maketitle

\begin{abstract}
Today, the security of many domains rely on the use of Machine Learning to detect threats, identify vulnerabilities, and safeguard systems from attacks. Recently, transformer architectures have improved the state-of-the-art performance on a wide range of tasks such as malware detection and network intrusion detection. But, before abandoning current approaches to transformers, it is crucial to understand their properties and implications on cybersecurity applications. In this paper, we evaluate the robustness of transformers to adversarial samples for system defenders (i.e., resiliency to adversarial perturbations generated on different types of architectures) and their adversarial strength for system attackers (i.e., transferability of adversarial samples generated by transformers to other target models). To that effect, we first fine-tune a set of pre-trained transformer, Convolutional Neural Network (CNN), and hybrid (an ensemble of transformer and CNN) models to solve different downstream image-based tasks. Then, we use an attack algorithm to craft 19,367 adversarial examples on each model for each task. The transferability of these adversarial examples is measured by evaluating each set on other models to determine which models offer more adversarial strength, and consequently, more robustness against these attacks. We find that the adversarial examples crafted on transformers offer the highest transferability rate (i.e., 25.7\% higher than the average) onto other models. Similarly, adversarial examples crafted on other models have the lowest rate of transferability (i.e., 56.7\% lower than the average) onto transformers. Our work emphasizes the importance of studying transformer architectures for attacking and defending models in security domains, and suggests using them as the primary architecture in transfer attack settings.
\end{abstract}


\section{Introduction}
Artificial intelligence, especially machine learning, is revolutionizing how we carry out cybersecurity\cite{ml_cybersecurity}. Understanding the capabilities and vulnerabilities of applying deep learning models in these domains from a practical standpoint is necessary. Emerging technologies in machine learning rely on transformer architectures~\cite{vaswani_attention_nodate}. Transformers provide the backbone of Large-Language Model's (LLM's): systems trained on massive amounts of data that can understand and generate contextually relevant text. While transformers have been known for state-of-the-art performance on Natural Language Processing (NLP)~\cite{vaswani_attention_nodate} and image classification~\cite{dosovitskiy_image_2021} tasks, their application has also improved cybersecurity domains such as malware detection~\cite{rahali2021malbert} and network intrusion detection~\cite{transformer_nids}.

With the incorporation of transformers in cybersecurity, ensuring the dependable operation of these systems in the presence of \textit{adversarial examples}, inputs with carefully generated perturbations to trigger misclassification, is vital. For example, adversarial malicious traffic records can be deliberately crafted to evade intrusion detection~\cite{lin_idsgan_2022}. There exists a large surface of attacks that systematically generate these examples and optimize for minimizing robust accuracy with respect to attack components~\cite{sheatsley_space_2022, carlini_towards_2017, papernot_practical_2017}. Researchers have historically analyzed white-box threat models where the adversary has complete access to model architecture and parameters. However, this threat model is unrealistic when models are deployed on systems where the adversary is unlikely to have access. Realistic adversaries will leverage \textit{transferability}: wherein adversarial examples crafted from one model are often adversarial examples for another \cite{papernot_practical_2017}. As transformer's take the spotlight, emerging as the latest leading architecture in performance, it is crucial that we understand their limitations when used for cybersecurity. 

Seminal papers have shown that transformers are implicitly more robust to white-box attacks~\cite{whitebox_transformer,mahmood_robustness_2021,bhojanapalli_understanding_2021,shao_adversarial_2022,paul_vision_2021,benz_adversarial_2021}. However, whether and how much this inherent robustness translates to black-box transfer attack settings with fine-tuned models remains unclear. In practice, there is an emergence and growing popularity of foundation models~\cite{bommasani_opportunities_nodate}(e.g., GPT-4~\cite{openai_gpt-4_2023} and DALL-E-2~\cite{ramesh_hierarchical_2022}), where transformers are pre-trained on massive general data and fine-tuned on a smaller dataset for target downstream tasks. The architecture becomes more versatile; leveraging the learned representations of the pre-training data to exploit the fine-tuning data. 

In this paper, we measure the adversarial strength and robustness of the generated adversarial examples from pre-trained transformer architectures with fine-tuning on a multitude of datasets. With transformer-based models at the forefront of image classification tasks, we take this domain as an instance for our study on the transformer's inherent security properties. We experiment using three sets of image classification models between transformer architectures (ViT~\cite{dosovitskiy_image_2021}), Convolutional Neural Networks (CNNs) (ResNet50 and ResNet152~\cite{he_deep_2015}), and hybrid model architectures that leverage both CNN and transformer architectures (DeiT, ConViT, and Swin~\cite{liu_swin_2021,liu_convnet_2022,touvron_training_2021}). Each model is then considered to craft adversarial examples using the Projected Gradient Descent (PGD)~\cite{madry_towards_2017} attack. The adversarial examples are subsequently fed to the other models in order to measure the resulting accuracy, thus the transferability, of these examples. Finally, we compare this approach across three different datasets to evaluate if transferability is further affected by the fine-tuning dataset. We find that the transformer provides more adversarial strength as an attacker in transferring adversarial examples crafted on it to other models, and more robustness as a defender against adversarial examples crafted on other models.

In this work, we contribute the following:
\begin{enumerate}
  \item We measure the transferability of adversarial examples to and from transformer architectures under practical scenarios where the model is fine-tuned on a dataset nonidentical from the pre-training.
  \item We demonstrate that transformers exhibit high transferability rates of adversarial examples onto other model architectures.
  \item We show that transformers offer more robustness in defense to adversarial examples crafted on other models.
\end{enumerate}

\section{Background}
\subsection{Architectures}
There are a variety of competitive architectures designed to perform image classification. Three of the most popular types of architectures include: CNNs, ViTs, and hybrid models. Following the introduction of convolutional techniques with LeNet~\cite{lecun_gradient-based_1998} in 1998, CNNs have maintained dominance as the primary architecture (e.g., ResNet~\cite{he_deep_2015} and VGG~\cite{simonyan_very_2015}) for a diverse range of computer vision tasks.

\textbf{Transformer and Hybrid Models}.
While transformers were principally designed to solve NLP tasks~\cite{vaswani_attention_nodate}, its versatility allows it to be applied to other domains such as cybersecurity~\cite{rahali2021malbert, transformer_nids} and image classification~\cite{dosovitskiy_image_2021}. The transformer architecture (ViT) relies on a self-attention mechanism which aims to capture \textit{global} relationships between words in NLP, or for instance, between image patches in the context of image classification. 

The introduction of transformer-based architectures in computer vision has led to the creation of hybrid models. A hybrid model is defined as a model combining the strength of CNNs and transformers, either architecturally or during training. We choose three of the existing state-of-the-art hybrid models: Swin Transformer~\cite{liu_swin_2021} which uses a shifted windowing scheme to compute hierarchical self-attention layers, ConViT~\cite{dascoli_convit_2022} which introduces gated positional self-attention to smoothly incorporate a convolutional inductive bias into a transformer, and DeiT~\cite{touvron_training_2021} which applies a teacher-student distillation to train a convolution-free transformer using a CNN as a teacher.

\subsection{Attack Surfaces}
In this work, the heuristic of a traditional transfer attack~\cite{papernot_practical_2017} is considered -- attacking unknown target models with adversarial examples crafted on source models (e.g., surrogate models). After building a surrogate model via querying the target model, the objective of an adversary is to create misclassification by perturbing the input as little as possible on the surrogate model so that the adversarial samples are misclassified by the surrogate model but are still perceptually similar to the vanilla input at inference time. Then, the crafted adversarial samples are \textit{transferred} to attack the target model. Given that the surrogate model is not necessarily equivalent to the target, evaluating how adversarial samples transfer across these architectures becomes the decisive factor of the attack's success. There are a plethora of attack algorithms to generate perturbations~\cite{goodfellow_explaining_2015, madry_towards_2017,croce_reliable_2020}. Projected Gradient Descent (PGD) is widely acknowledged as one of the state-of-the-art methods for crafting adversarial examples, and thus we choose it to be the attack algorithm in this paper. The optimization objective can be written as:
\begin{gather}
    x^{t+1} = \prod_{x+\mathcal{B}} (x^t+\alpha\text{sgn}(\nabla_xL(\theta,x,y)))
    \label{eq-pgd}
\end{gather}
where $L$ is the loss function of the source model with parameters $\theta$, original input $x$, and label $y$. PGD applies gradient descent steps $\alpha$ iteratively on the input space (i.e., $x^t$ and  $x^{t+1}$) while ensuring that the perturbed input lies within a predefined vicinity of the original input measured by $\ell_\infty$-norm ball $\mathcal{B}$. This metric aims to craft adversarial examples that are both effective (i.e., misclassified by the models) and feasible (i.e., imperceptible to humans).

\subsection{Robustness and Transferability}
Many works~\cite{garg_bae_2020,goyal_survey_2023} in NLP have shown that transformers have higher robustness compared to CNNs against attacks. Despite the shift from discrete perturbations in NLP domain to continuous perturbations in computer vision, transformer-based architectures continue to demonstrate stronger resiliency against adversarial examples than CNN-based models ~\cite{bhojanapalli_understanding_2021,shao_adversarial_2022,paul_vision_2021,benz_adversarial_2021}. While their definition of robustness refers to the model's accuracy when confronted with adversarial examples generated by the same architecture, a more pragmatic question arises: \textit{In scenarios where adversaries do not have white-box access to target models, do transformer-based models still exhibit higher robustness against adversarial samples crafted by various models that may not necessarily share the same architecture, weights, and training recipe?} Previous work~\cite{shao_adversarial_2022,mahmood_robustness_2021} has identified a limited transferability of adversarial examples between CNNs and transformers. However, these studies lack a comprehensive analysis of the varying degrees of efficacy in the two directions of transfer attack -- adversarial strength and robustness. In addition, while comparing model error rates, they measure transferability on adversarial examples generated using the entire dataset. This does not take account of the original accuracy (i.e., on vanilla images) of the models, resulting in a distortion of the measure of transferability. Furthermore, to the best of our knowledge, no existing work has delved into the transferability of attacks on models fine-tuned on different \textit{downstream tasks}; a widely adopted application of large foundation models in practice.

\section{Transferability Study}
\begin{figure*}[!ht]
    \centering
    \includegraphics[width=0.8\textwidth]{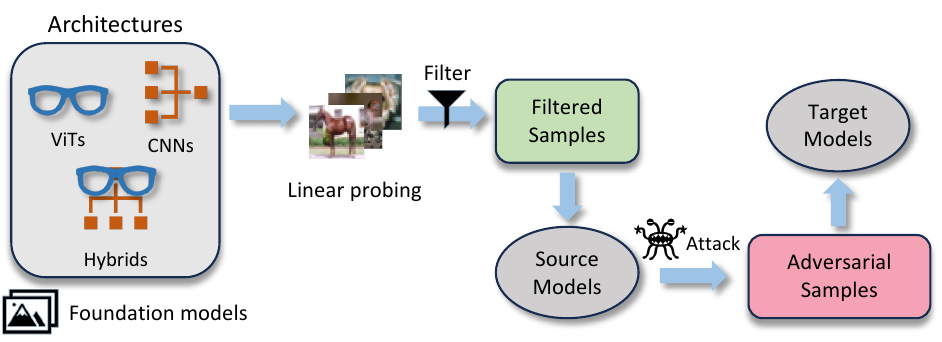}
    \caption{\textbf{Methodology Overview}: We fine-tune a set of pre-trained models with various architectures on different downstream tasks. Then, we filter the input samples (i.e., only correctly classified samples are kept) and craft adversarial examples on them with each model. Finally, we evaluate the transferability (i.e., \textit{adversarial strength} and \textit{robustness}) of different architectures.}
    \label{fig:methdology_figure}
\end{figure*}

In this paper, we consider the threat model to be a black-box scenario: the target model is deployed on systems where the adversary has no access (i.e., no knowledge on model architecture and weights). Thus, our evaluation is done by crafting adversarial examples on a surrogate model with an arbitrary architecture (CNN, transformer, hybrid). It is assumed that the adversary is aware that the model has undergone pre-training on a common massive dataset and is applied to solve a specific downstream task. The framework (shown in \autoref{fig:methdology_figure}) for measuring transferability of adversarial examples to and from architectures is instantiated for each dataset considered.

\subsection{Training}
In order to study the models in the same setting while considering multiple datasets, the following protocol is applied: the models are fine-tuned on the same dataset given that they are all pre-trained on the same large dataset. This fine-tuning approach follows how a foundation model would be used for a downstream task in practice. This helps us answer whether fine-tuning pre-trained models will preserve the model's ability to generate effective adversarial examples and its resilience against them. 

Six models from PyTorch Image Models (TIMM)~\cite{wightman_rwightmanpytorch-image-models_2023} pre-trained on ImageNet-1K~\cite{deng_imagenet_2009} with the same training image size, i.e., 224x224 are studied:
\begin{itemize}
    \item 2 CNN-based ResNet~\cite{he_deep_2015} models of 50 layers (25.6M parameters) and 152 layers (60.2M parameters) respectively.
    \item 1 transformer-based ViT model~\cite{dosovitskiy_image_2021} (86.6M parameters).
    \item 3 hybrid models: DeiT~\cite{touvron_training_2021}, ConViT~\cite{dascoli_convit_2022}, and Swin Transformer~\cite{liu_swin_2021} (86.6M, 86.5M, and 71.1M parameters).
\end{itemize}

To preserve the existing trained representation from pre-training, a linear probing approach is considered. This prevents the models from modifying all weights as seen in traditional fine-tuning, which is computationally expensive. Linear probing aims to train a classifier while leveraging the pre-trained representation of input images and the labels from downstream training datasets.

More specifically, our evaluation uses the following hyperparameters: learning rate $\eta=0.001$, SGD momentum $m=0.9$, number of epochs $N_{e}=100$, and batch size $b=100$, and the loss considered is the cross-entropy loss. Moreover, all images are resized to 224x224 and normalized with a mean and standard deviation given by each pre-trained model's configuration during training and attacks. 

The three datasets used for fine-tuning are the following:
\begin{itemize}
    \item CIFAR-10~\cite{krizhevsky_learning_nodate} made of 60,000 32x32 color images in 10 classes, with 6,000 images per class.
    \item CIFAR-100~\cite{krizhevsky_learning_nodate} made of 60,000 32x32 color images in 100 classes, with 600 images per class. 
    \item STL-10~\cite{coates_analysis_nodate}, a subset of ImageNet made of 13,000 96x96 color images in 10 classes, with 1,300 images per class.
\end{itemize}

\subsection{Attacking}
After linear probing, the samples from the test set that can be correctly classified by \textit{all} models are identified. This resulting subset of the test set is then used to craft the adversarial examples. This filtering is made in order to accurately measure the vulnerability of models to adversarial examples. This set, referred to as \textit{filtered samples}, serves as targets for attacks. Then, PGD is used to craft adversarial examples from the 6 fine-tuned models for each downstream tasks to then be analyzed in their transfer attack success rate when fed to other models.

The number of filtered images of each dataset is dependent on the accuracy of models on them. After filtering the test set (originally 10k images for CIFAR10 and CIFAR100, and 8k images for STL10), CIFAR10, CIFAR100, and STL10 now have 7,308, 4,542, and 7,517 filtered samples, respectively.
With the PGD algorithm and 6 linear-probed models per dataset, $(7,308+4,542+7,517) * 6 = 116,202$ adversarial examples are crafted in total. More specifically, the perturbation budget $\epsilon$ is set to $\frac{4}{255}$\footnote{This choice is made in order to show separation among the results. Notably, at $\frac{8}{255}$, many models exhibit a 100\% transferability rate.}, $\alpha=\frac{2}{255}$, and steps=$10$ for the attack. These hyperparameters, including the budget, are fixed across models and datasets for a fair comparison.

\subsection{Transferring}\label{sec:transfer}
Given a set of source and target models $(s,t)\in \mathcal{M}^{2}$, a set of adversarial examples $\mathcal{A}_{s}^{(d)} = \{(x,y)\}$ are crafted using PGD on the source model $s$ with a dataset $d\in\mathcal{D}$, it is possible to define a transferability score $\mathcal{T}^{(d)}(s,t)$ as the percentage of adversarial examples from $\mathcal{A}_{s}^{(d)}$ misclassified by the target model $t$ which can be formalized as:
\begin{equation}\label{eq:transferability_score}
\mathcal{T}^{(d)}(s,t) = \frac{|\{(x,y)\in\mathcal{A}_{s}^{(d)}; \hat{y}_{t}\neq y\}|}{|\mathcal{A}_{s}^{(d)}|}    
\end{equation}

Transferability is then broken down into two components that we coin as transmissibility (i.e., an indicator of adversarial strength) and receivability (i.e., an indicator of robustness). For a given model $m\in\mathcal{M}$ and a given dataset $d\in\mathcal{D}$, the transmissibility $\mathcal{T}_{t}^{(d)}(m)$ is defined as the Euclidean norm of the vector made of the transferability scores with $m$ as the source model. Similarly, the receivability $\mathcal{T}_{r}^{(d)}(m)$ corresponds to the Euclidean norm of the vector made of the transferability scores with $m$ as the target model. This can be summarized as follows:
\begin{equation}
   \mathcal{T}_{t}^{(d)}(m) = \lVert(\mathcal{T}^{(d)}(m,t))_{t\in\mathcal{M}, t\neq s}\rVert_{2}
\end{equation}
\begin{equation}
    \mathcal{T}_{r}^{(d)}(m) = \lVert(\mathcal{T}^{(d)}(s,m))_{s\in\mathcal{M}, s\neq t}\rVert_{2}
\end{equation}
As such, the transmissibility $\mathcal{T}_{t}^{(d)}(m)$ measures how well adversarial examples crafted on $m$ transfer to other models. The receivability $\mathcal{T}_{r}^{(d)}(m)$, on the other hand, quantifies how well examples crafted from other models transfer to $m$. Recognizing this distinction provides insights to both the attacker and the defender. The former aims to maximize transmissibility in hopes to increase the likelihood of adversarial examples being successfully transferred to the target model. Likewise, the latter aims to minimize receivability in hopes to decrease the chances of adversarial examples being effectively transferred to the model.

\subsection{Results}
\begin{figure}
     \centering
     \begin{subfigure}[b]{0.45\textwidth}
         \centering
         \includegraphics[width=\textwidth]{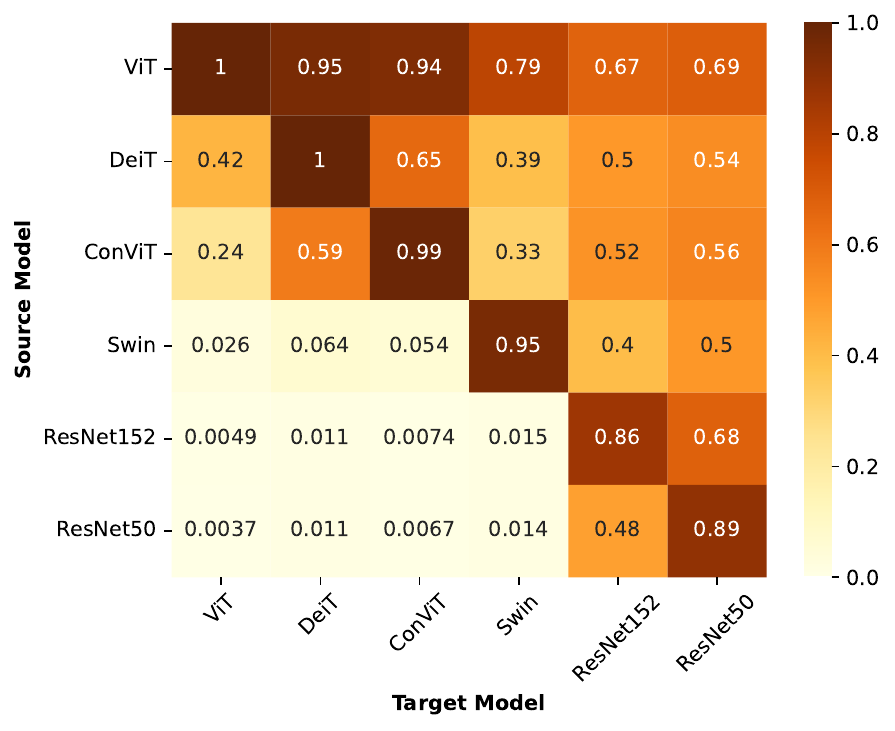}
         \caption{CIFAR-10}
         \label{fig:cifar10_heatmap}
     \end{subfigure}
     \hfill
     \begin{subfigure}[b]{0.45\textwidth}
         \centering
         \includegraphics[width=\textwidth]{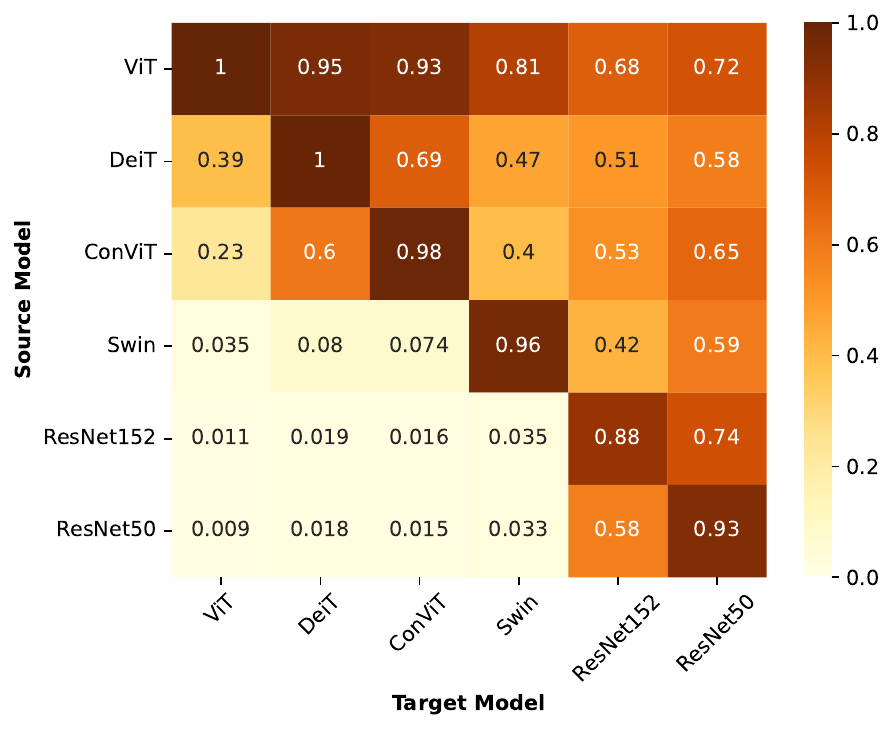}
         \caption{CIFAR-100}
         \label{fig:cifar100_heatmap}
     \end{subfigure}
     \hfill
     \begin{subfigure}[b]{0.45\textwidth}
         \centering
         \includegraphics[width=\textwidth]{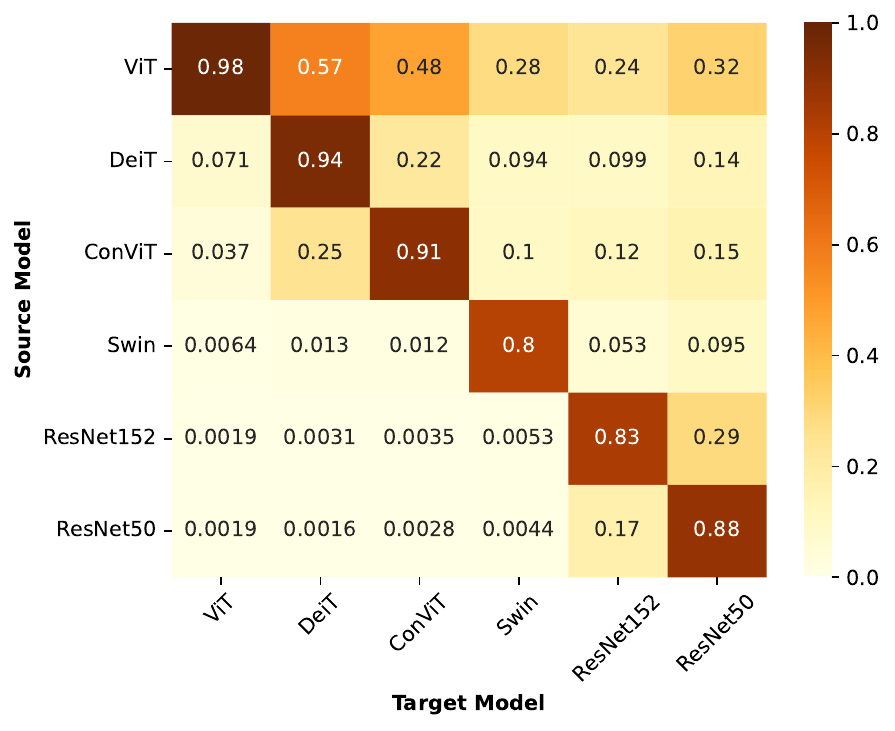}
         \caption{STL-10}
         \label{fig:stl10_heatmap}
     \end{subfigure}
        \caption{\textbf{Heatmaps of Transferability}: Entries represent the transferability score $\mathcal{T}^{(d)}(s,t)$ of Source Model $s$ to Target Model $t$ on each dataset $d$ denoted in each subfigure.}
        \label{fig:heatmaps}
\end{figure}
\textbf{Heatmap Evaluation.}
For each dataset, the adversarial examples crafted by the 6 models are used to cross-attack each other, resulting in 36 transfer attack experiments. Using \autoref{eq:transferability_score}, the transferability scores are calculated for every experiment, resulting in three heatmaps for CIFAR10 (\autoref{fig:cifar10_heatmap}), CIFAR100 (\autoref{fig:cifar100_heatmap}), and STL10 (\autoref{fig:stl10_heatmap}).

Each heatmap is constructed with the source models spanning the rows and the target models spanning the columns. In order to study the influence of either of the two architectures, the models are ordered from ViT to CNN, with hybrid models in the middle. Each entry in the heatmap refers to the percent of adversarial examples misclassified by the target model, as defined by $\mathcal{T}^{(d)}(s,t)$ in \autoref{eq:transferability_score}. 

In \autoref{fig:cifar10_heatmap}, the heatmap is largely upper triangular on the CIFAR-10 dataset. This means that (1) the transferability of adversarial examples decreases when reading source models from top to bottom, and (2) the transferability of adversarial examples increases when reading target models from left to right. More specifically, as the source model gets closer to a CNN-like architecture, the transfer score diminishes. Likewise, as the target model becomes more related to a CNN compared to a ViT, the transfer score becomes higher. The same trend can be observed in \autoref{fig:cifar100_heatmap} on the CIFAR-100 dataset. Observing \autoref{fig:stl10_heatmap}, it is evident that the same magnitude of upper triangularity from CIFAR-10 and CIFAR-100 is not visible on the STL-10 dataset. However, the same general trend of source and target model transferability from CIFAR-10 and CIFAR-100 exists in a much more subtle view.
\begin{figure}[t]
    \centering
    \includegraphics[width=0.4\textwidth]{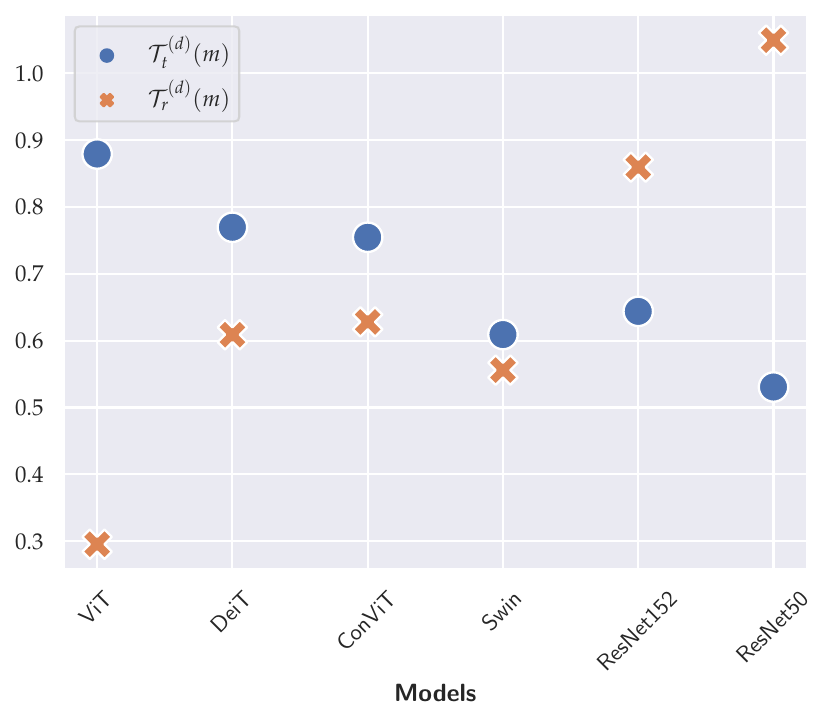}
    \caption{\textbf{Transmission and Reception of Adversarial Examples}: This figure illustrates the normalized transmissibility $\mathcal{T}_{t}^{(d)}(m)$ and receivability $\mathcal{T}_{r}^{(d)}(m)$ rates to and from each model on CIFAR-100.}
    \label{fig:attack_defense}
\end{figure}

\textbf{Transmissibility and Receivability Evaluation.}
To formalize the transferability of adversarial examples from source models and to target models, we refer to the transmissibility $\mathcal{T}_{t}^{(d)}(m)$ and receivability $\mathcal{T}_{r}^{(d)}(m)$ defined in \autoref{sec:transfer}. Under these metrics, a model provides more adversarial strength to an attacker if it has a high transmissibility, and more robustness to a defender if it has a low receivability. \autoref{fig:attack_defense} illustrates these scores across each model on CIFAR-100. Similar to the heatmaps, the models are ordered from ViT, hybrids, to CNN to illustrate the middle-ground of ViT's and CNN's. 

From the transmissibility scores $\mathcal{T}_{t}^{(d)}(m)$, we observe a decreasing trend as the architecture becomes more similar to a CNN rather than a ViT; an observation made on the heatmaps when viewing source models top to bottom. Oppositely, an increasing trend can be observed on the receivability scores $\mathcal{T}_{r}^{(d)}(m)$; an observation also made on the heatmaps when reading target models left to right. Although these scores are only given on the CIFAR-100 dataset, it has to be noted that the same trend across source and target models in CIFAR-10 and STL-10 was present. That is, the transmissibility of source models decreasing and receivability of target models increasing as architectures became more CNN-like from the ViT. 

Indeed, the ViT exhibits the highest transmissibility score of $0.88$ and lowest receivability score of $0.30$ on CIFAR-100. Consequently, the CNN architectures (ResNet152 and ResNet50) provide the lowest average transmissibility score of $0.59$ and highest average receivability score of $0.96$. This shows that CNN architectures provide attackers little strength and defenders a vulnerable model. Interestingly, the hybrid architectures (DeiT, ConViT, and Swin) lie in between the transmissibility and receivability scores of the transformer and CNN architectures with an average transmissibility score of $0.71$ and average receivability score of $0.60$. Thus, they do not leverage the benefits of both transformer and CNN architectures to make an overall stronger adversary or more resilient defender. This means that the transformer architecture itself provides an attacker more adversarial strength, and a defender more robustness under this transfer attack setting.

\subsection{On Cybersecurity}
Given the depth and maturity of research on model architectures and the corresponding adversarial example generation algorithms within the realm of computer vision, this study has principally centered its experiments on the image domain. This strategic focus is aimed at enabling a comprehensive comparison on the transferability of adversarial examples among transformers, CNNs, and hybrid models. Despite the disparity in inputs -- such as source code of an Android application~\cite{rahali2021malbert} versus images -- the utilization of transformers in both malware detection and image classification involves common underlying implementation blocks. Furthermore, as previously highlighted in the results section, the transferability of attacks exhibits considerable variability across distinct model architectures but little dependence on input datasets. This characteristic represents a potential transition from computer vision security to other cybersecurity domains. While whether the findings of this study on images can be generalized to other specific domains (e.g., malware detection and network intrusion detection) or not still needs more future work, the approaches and algorithms in our transferability framework set up a stepping stone for both future large-scale implementation and evaluation. To summarize, with the anticipation of cybersecurity adopting transformer architectures, our study offers practical recommendations and caveats to practitioners in the field for selecting and adapting models for enhancing their resilience against transfer-based adversarial attacks. 

\section{Conclusion}
In this paper, we investigated the security implications of the transformer architecture with respect to the transferability of adversarial examples. While this architecture is growing in popularity in cybersecurity applications, too little is known on its robustness and abilities in adversarial settings. The transferability of adversarial examples was measured through 108 experiments across different models pre-trained and adapted to a downstream task. More precisely, two properties were studied: how well adversarial examples crafted from a model transfer, and how robust a model is against adversarial examples crafted on other architectures. The results show that transformer architectures offer the best adversarial strength and robustness compared to similar architectures, regardless of the fine-tuning dataset. This shows that the transformer architecture should be the de facto model when defending against transfer attacks in black-box settings compared to other architectures. Additional results show hybrid models not having as much adversarial strength and robustness as transformers in a transfer attack setting. This suggests more work to be done in explaining the possible trade-off between leveraging different architectures for performance and risking vulnerability to transfer attacks. Moreover, given the number of parameters influencing transferability (e.g., attack performed and perturbation budget), a more thorough study could lead to a more systematic approach to transferability.

\section*{Acknowledgement}
\addcontentsline{toc}{section}{Acknowledgment}
We would like to thank Ryan Sheatsley, Blaine Hoak, Yohan Beugin, Eric Pauley, and Quinn Burke for their helpful comments on earlier versions of this paper.

\textbf{Funding Acknowledgment:} This material is based upon work supported by, or in part by, the Combat Capabilities Development Command Army Research Laboratory under Grant No. W911NF-21-1-0317, the National Science Foundation under Grant No. CNS-2343611, and the Semiconductor Research Corporation (SRC) and DARPA under Grant No. 705776. Any opinions, findings, and conclusions or recommendations expressed in this publication are those of the author(s) and do not necessarily reflect the views of the Combat Capabilities Development Command Army Research Laboratory, National Science Foundation, or SRC and DARPA. The U.S. Government is authorized to reproduce and distribute reprints for government purposes notwithstanding any copyright notation hereon. 

\bibliographystyle{IEEEtran}
\bibliography{ref}

\end{document}